\documentclass[aps,twocolumn,amsmath,amssymb,
showkeys,floatfix]{revtex4}
\usepackage{graphicx}
\usepackage{amssymb}
\usepackage{color}
\usepackage{float}
\usepackage{mathrsfs}
\usepackage{tabularx}

\begin{document}



\title{Pre-equilibrium effects in charge-asymmetric low-energy reactions} 

\author{H. Zheng$^{1}$\footnote{Email address: zheng@lns.infn.it}, S. Burrello$^{1,2}$,  M. Colonna$^{1}$, V. Baran$^{3}$ }
\affiliation{$^{1}$ Laboratori Nazionali del Sud, INFN, I-95123 Catania, Italy}
\affiliation{$^{2}$ Physics and Astronomy Department, University of Catania, Italy}
\affiliation{$^{3}$ Faculty of Physics, University of Bucharest, Romania}


\begin{abstract}
We study the pre-equilibrium dipole 
response in the charge-asymmetric reaction  $^{132}$Sn+$^{58}$Ni at $E_{lab}=10$ MeV/A, within a semi-classical transport
model employing effective interactions for the nuclear mean-field. 
In particular, we adopt the recently introduced
SAMi-J Skyrme interactions, whose parameters are specifically tuned to
improve the description of spin-isospin properties of nuclei. Within the same framework, we also discuss
pre-equilibrium nucleon emission.
Our results show that 
both mechanisms, i.e., pre-equilibrium dipole oscillations and nucleon 
emission,   
are sensitive to the symmetry energy below the saturation density $\rho_0$ 
(in the range $0.6\rho_0-\rho_0$),
to the effective mass and to the nucleon-nucleon cross section.  
Finally, a covariant analysis is applied to study the correlations between the model parameters and observables
of experimental interest.

\end{abstract}



\maketitle


\section{Introduction}

Heavy ion reactions at energies just above the Coulomb barrier are 
governed, to a large extent, by one-body dissipation mechanisms.
The main reaction path ranges from (incomplete) fusion to binary exit  
channels, such as deep-inelastic or quasi-fission processes. However,
in spite of the apparent simplicity of the reaction dynamics,
quite intriguing features may manifest along the fusion/fission path, reflecting
the complexity of the self-consistent mean-field, and structure effects may still play a relevant role \cite{Washi09,Umar16}.

Furthermore, new interesting 
phenomena, linked to the charge equilibration mechanism, appear in reactions between charge-asymmetric systems.  
If the reaction partners have appreciably different
N/Z ratios, the proton and neutron centers of mass of the reacting
system may not coincide during the early stages of the
collision. Then, 
apart from incoherent exchange of nucleons
between the colliding ions,  
collective oscillations of protons against
neutrons might occur on the way to fusion, along the symmetry
axis of the composite system. 
This is the so-called
dynamical dipole (DD) mode or pre-equilibrium 
giant dipole resonance (GDR) \cite{baranPRL2001,simenelPRL2001,Umar07,
wuPRC2010,flibottePRL1996,Papa05,pierrPRC2009,Giaz14}. 
In (incomplete) fusion
reactions, the shape of the pre-equilibrium dinuclear complex
exhibits a very large prolate deformation as compared to
the shape finally reached by the equilibrated compound nucleus.
Consequently, the pre-equilibrium radiation carries out 
relevant information 
about the shape of the system,  
 as well as insight into the charge equilibration,
and provides a cooling effect along the
fusion path, possibly favoring the formation of superheavy
elements \cite{simenelPRC2007,parasPRC2016}.
Thus one expects the DD to be influenced by different parameters,
like  charge and mass asymmetry, collision centrality and energy 
\cite{baranPRL2001,tson2001,pierrPRC2009}. 
One should also consider that, within the optimal beam energy 
for the investigation of the DD (around 10 MeV/A), 
pre-equilibrium emission of nucleons and light particles
can occur. This effect contributes to cool down the system
and may reduce the initial charge asymmetry of the colliding nuclei,
 owing to the favored neutron emission
in neutron-rich systems.


Similar to the GDR, DD oscillations are 
governed mainly by the isovector channel of the nuclear
effective interaction, which provides the restoring force.
We note that the isovector terms are connected to the symmetry energy of
the nuclear Equation of State (EoS),  on which several investigations
are concentrated nowadays \cite{barPR2005,li08,giuliani14}.
Indeed, the DD mechanism, as well as the N/Z ratio of the pre-equilibrium
nucleon emission, has been proposed as a possible tool
to probe the low-density behavior of the symmetry 
energy \cite{pierrPRC2009,Giaz14,baranPRC2009}. 
It is worthwhile to mention that the latter
plays an essential
role in nuclear structure \cite{piek12} (determining for instance the thickness of the
neutron skin in neutron-rich nuclei), as well as in the astrophysical
context, for the description of low-density clustering in 
compact stars \cite{steiner05,burrello15}.

As already stressed above, in low energy collisions, we expect 
structure effects to influence the reaction dynamics. To take into account this aspect, 
making a bridge between structure and reaction studies,
we will explore the charge equilibration dynamics also
employing new effective interactions, which are especially devised to
reproduce the properties
of nuclei (ground state and collective excitations,
particularly in the spin-isospin channels) \cite{coll4}.
One should also consider that, in the energy range of our interest, 
two-body correlations, beyond the mean-field picture, start to play 
a significant role.
The latter are described in terms of hard nucleon-nucleon (n-n) scattering,
representing the effect of the hard core on the nuclear interaction.
The goal of this Letter is to investigate, within a semi-classical
transport approach,  pre-equilibrium dipole radiation and nucleon emission in low-energy 
nuclear reactions, 
to get a deeper insight into 
their sensitivity to specific features of
the nuclear effective interaction and 
n-n cross section (cs), in the density range explored along the reaction dynamics.   

\begin{figure}
\begin{center}
\includegraphics*[scale=0.36]{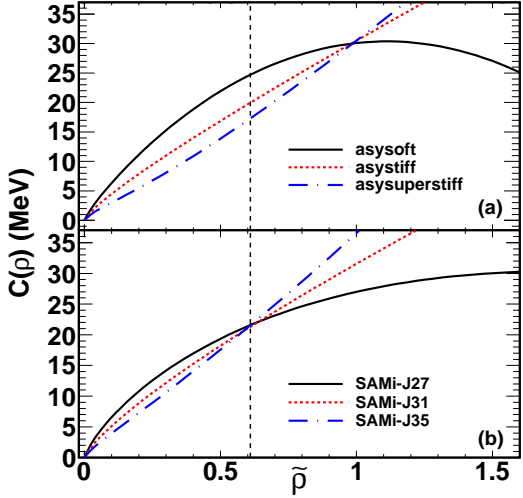}
\end{center}
\caption{(Color online) The symmetry energy versus reduced density $\tilde{\rho} = \rho/\rho_0$ for the EoS without (a) and with (b) momentum dependence. The vertical dashed line is to guide the eye to the cross point of the three SAMi-J EoS.}
\label{eossym}
\end{figure}

\section{The model description}

Calculations will be performed employing the Boltzmann-Nordheim-Vlasov 
(BNV) model \cite{Bon94,Guar96}. 
Essentially, one solves 
the two coupled kinetic equations for the neutron and proton
 distribution functions $f_q({\bf r},{\bf p},t)$, with $q=n,p$, respectively 
\cite{barPR2005}:
\begin{equation}
\frac{\partial f_q({\bf r},{\bf p},t)}{\partial t}+\frac{\partial \epsilon_q}{\partial {\bf p}}\frac{\partial f_q ({\bf r},{\bf p},t)}{\partial {\bf r}}-
\frac{\partial \epsilon_q}{\partial {\bf r}}\frac{\partial f_q({\bf r},{\bf p},t)}{\partial {\bf p}}= I_{coll}[f_n,f_p] ,
\label{vlasov}
\end{equation}
where  $\epsilon_q$ represents the single particle energy, which can be deduced from the
energy density, $\mathscr{E}$ \cite{Lar98}. 
Considering a standard Skyrme interaction, the latter is expressed in terms of the isoscalar, $\rho=\rho_n+\rho_p$,
and isovector, $\rho_{3}=\rho_n-\rho_p$,  densities and 
kinetic energy densities ($\tau=\tau_{n}+\tau_{p}, \tau_{3}=\tau_{n}-\tau_{p}$) as \cite{radutaEJPA2014}:
\begin{eqnarray}
\mathscr{E}&=&\frac{\hbar^2}{2 m}\tau + C_0\rho^2 + D_0\rho_{3}^2 + C_3\rho^{\alpha + 2} + D_3\rho^{\alpha}\rho_{3}^2 ~+ C_{eff}\rho\tau \nonumber\\
&& + D_{eff}\rho_{3}\tau_{3} + C_{surf}(\bigtriangledown\rho)^2 + D_{surf}(\bigtriangledown\rho_3)^2,
\label{eq:rhoE}
\end{eqnarray}
where the coefficients $C_{..}$, $D_{..}$ are combinations of traditional Skyrme parameters. 
The Coulomb interaction is also considered in the calculations \cite{zhengPRC2016}.

In the collision integral, $I_{coll}[f_n,f_p]$,
which accounts for residual two-body correlations,  
we employ the
energy, angular and isospin dependent free n-n cross section \cite{Bar02}.  
The integration of Eq. (\ref{vlasov}) is based on the test-particle (t.p.) method \cite{wong}. 

While the model can not account for effects associated with the shell structure, this self-consistent approach is able to describe robust quantum modes, of zero-sound type,
in both nuclear matter and finite nuclei \cite{barPR2005, urbPRC2012}, and also low-energy reaction dynamics \cite{Rizzo14}.

We are mostly interested in the effects linked to the isovector terms of the nuclear effective interaction, thus we introduce 
the symmetry energy per nucleon, $E_{sym}/ A = C(\rho) I^2$, 
where $I = \rho_3/\rho$ is the asymmetry parameter and the coefficient $C(\rho)$ can be written as a function of the Skyrme coefficients (at zero temperature):
\begin{equation}
C(\rho) = \frac{\varepsilon_F}{3} + D_0\rho + D_3\rho^{\alpha+1} ~+ 
\frac{2m}{\hbar^2}\left(\frac{C_{eff}}{3} + D_{eff}\right)\varepsilon_F\rho,
\end{equation}
with $\varepsilon_F$ denoting the Fermi energy at density $\rho$.  


In the following we will adopt the recently introduced SAMi-J Skyrme effective interactions \cite{coll4}. 
The corresponding parameters have been fitted based on the SAMi fitting protocol \cite{coll4}: binding energies and charge radii of some doubly magic nuclei, which allow the SAMi-J family to predict a reasonable saturation density ($\rho_0=0.159$ fm$^{-3}$), energy per nucleon $E/A (\rho_0) =-15.9$ MeV and incompressibility modulus ($K = 245$ MeV) of symmetric nuclear matter; some selected spin-isospin sensitive Landau-Migdal parameters \cite{caoPRC2010};  the neutron 
matter EoS of Ref.\cite{wiringaPRC1988}. 
Moreover, these interactions lead to 
an effective isoscalar nucleon mass $m^*(I = 0) = 0.67~m$ and a
neutron-proton effective mass splitting 
$m^*_n - m^*_p = 0.023~mI$ MeV at saturation density (being $m$ the nucleon mass).
In this study, we will test these interactions for the description of reaction
mechanisms, 
employing, in particular, three SAMi-J parametrizations: SAMi-J27, SAMi-J31 and SAMi-J35 \cite{coll4}. 
Since, as mentioned above, the SAMi-J interactions have been fitted in order to also reproduce the main features of selected nuclei, 
the symmetry energy coefficient takes the same value, $C(\rho_c) \approx 22$ MeV, 
at the density $\rho_c \approx 0.6\rho_0$, 
which can be considered as the average density of medium-size nuclei. 
Then the value, J, of the symmetry energy at saturation density is different in the three cases, being equal to 27 MeV (SAMi-J27), 31 MeV (SAMi-J31)
and 35 MeV (SAMi-J35), respectively. The values of the slope parameter
 $\displaystyle L = 3 \left. \rho_0 \frac{d C(\rho)}{d \rho} \right\vert_{\rho=\rho_0}$ are equal to $L = 29.9$ MeV (SAMi-J27), $L = 74.5$ MeV (SAMi-J31)  and $L = 115.2$ MeV (SAMi-J35).  
The density dependence of $C(\rho)$ is shown in Fig. \ref{eossym} (panel (b)). 
\begin{figure}
\begin{center}
\includegraphics*[scale=0.36]{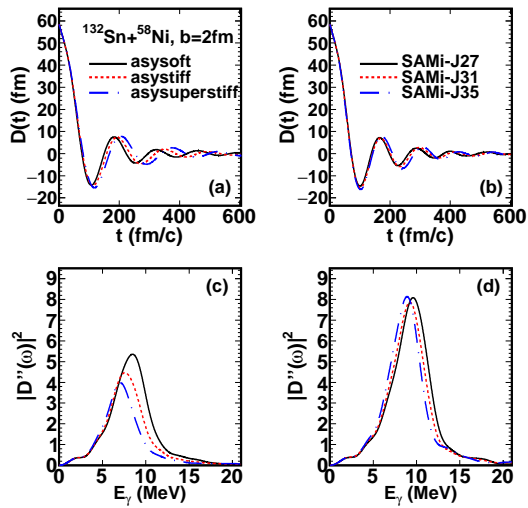}
\end{center}
\caption{(Color online) (top) The time evolution of DD for the MI and MD EoS at b=2 fm; (bottom) the corresponding power spectrum of the dipole acceleration (see text).}
\label{ddeos}
\end{figure} 

In order to compare with previous studies \cite{baranPRC2009, coll2,zhengPRC2016}, we shall also consider simplified Skyrme interactions without momentum dependence ($C_{eff} = D_{eff} = 0$, 
$m^* = m$), corresponding to an incompressibility modulus equal to $K = 200$ MeV \cite{coll2}.  
We will refer to these interactions as momentum-independent (MI) interactions, 
to be distinguished from the  momentum dependent (MD) SAMi-J 
parametrizations. 
As far as the symmetry energy is concerned, 
the MI parametrizations 
also allow for three different types of density dependence \cite{coll2}, which
are characterized by a very similar value of J ($\approx$ 30 MeV), but a different slope 
parameter $L$: $L = 14.8$ MeV (asysoft),  $L = 79$ MeV (asystiff) and
$L=106$ MeV (asysuperstiff).
Thus, as one can observe in Fig. \ref{eossym} (panel (a)), the three parametrizations of the symmetry energy cross each other at 
$\rho = \rho_0$ in this case. 
We will see in the following that the use of such a variety of effective interactions allows one to better identify the density region probed by the
reaction mechanisms considered in our study.

\section{Results and discussions}

BNV calculations,  employing both MI and MD interactions, have been carried out for
the reaction $^{132}$Sn+$^{58}$Ni at 10 MeV/A, considering 
different impact parameters,  b=0, 2, 4 and 6 fm, which lead to incomplete
fusion. 
A number of 600 t.p. per nucleon was used in all cases, to ensure a good spanning of the distribution function in phase space. 
Reactions involving the very neutron rich  $^{132}$Sn allow one to
inject a significant charge asymmetry in the entrance channel and, possibly,
to explore neutron-skin effects.
Since the ratios $(\frac{N}{Z})_P=1.64$ and $(\frac{N}{Z})_T=1.07$ are different for projectile (P) and target (T), 
a sizeable isovector dipole moment is excited in the initial conditions, which can trigger a DD oscillations along the rotating reaction symmetry axis. 
One has also to consider that, at the studied beam energy, pre-equilibrium nucleon emission starts to play a significant role. 
Thus it is worthwhile to investigate on equal footing both pre-equilibrium effects, i.e., particle and $\gamma$-ray emission.


Let us start by discussing dipole oscillations. 
As done in previous studies investigating the DD $\gamma$ decay \cite{baranPRL2001, baranPRC2009, parasPRC2016}, we adopt a collective bremsstrahlung analysis. 
The dipole moment, in coordinate
space, is defined as: 
\begin{equation}
D(t)=\frac{NZ}{A}(R_p-R_n),
\end{equation}
where $A=A_P+A_T$ is the total mass of the dinuclear system and $N=N_P+N_T$ ($Z=Z_P+Z_T$) is the neutron (proton) number.
$R_p$ and $R_n$ refer to the center of mass of protons and neutrons, respectively. 
For the system considered, at the contact time between the two colliding nuclei, 
the dipole moment amounts to $D_0$ = 45.1 fm. 
Denoting by $E_\gamma$ the photon energy, 
the emission probability associated with dipole oscillations is given by ($E_\gamma=\hbar \omega$):
\begin{equation}
\frac{dP}{dE_\gamma} = \frac{2e^2}{3\pi \hbar c^3 E_\gamma} |D''(\omega)|^2, \label{gmdecay}
\end{equation}
where $D''(\omega)$ 
is the Fourier transform of the dipole acceleration $D''(t)$ 
\cite{baranPRL2001}. 
In order to reduce the numerical noise, 10 events have been considered for each impact parameter and for each
effective interaction employed, and the results shown in the following 
are obtained from the average DD evolution. 

\begin{figure}
\begin{center}
\includegraphics*[scale=0.36]{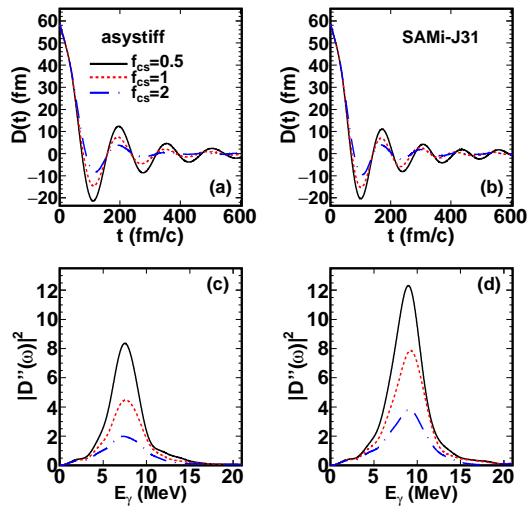}
\end{center}
    \caption{(Color online) Similar to Fig. \ref{ddeos}. Left panels are for the asystiff (MI) interaction and right panels are for SAMi-J31 (MD). 
Results are plotted for different choices of the n-n cross section (see text).
}
\label{ddcs}
\end{figure}

In Fig. \ref{ddeos} (a) and (b), we represent the time evolution of $D(t)$,
at b= 2 fm, for the MI and MD EoS, respectively. 
The initial time corresponds to a distance of 14 fm between the centers of mass
of the two nuclei. 
One observes damped oscillations, which are exhausted 
within about 600 fm/c. 
This behavior can be ascribed to both mean-field and two-body collisional 
damping effects.  
The corresponding energy spectrum,  $|D''(\omega)|^2$
, which, as shown by Eq. (\ref{gmdecay}), 
determines the $\gamma$ emission probability,  is also displayed 
in Fig. \ref{ddeos} (see (c) and (d) panels). 

One can see that there are several differences between the power spectra 
obtained for the MI and the MD EoS. 
First, in the MI case
 the centroids of the power spectra exhibit a dependence on the EoS, whereas they are close to each other for the MD EoS.
At the same time, one observes that there is a magnitude ordering of the power spectra for MI EoS, which is consistent with our previous studies \cite{baranPRC2009}, 
whereas the results are similar for the three MD EoS. 
Since the restoring force of isovector dipole oscillations is essentially provided by the symmetry energy, these findings suggest that 
DD probes 
the density region around the
crossing point of the SAMi-J interactions ($\rho \approx 0.6~\rho_0$). 
Indeed, whereas the
three SAMi-J (MD) parametrizations 
lead to quite
similar results, 
in the MI case the frequency and the magnitude of the power spectrum are higher in the soft case, reflecting the larger value of the
symmetry energy (see Fig. \ref{eossym}, panel (a)).
The sensitivity to this density domain reflects the elongated shape of 
the system during the pre-equilibrium phase, enhancing surface contributions.    

Second, 
the power spectra have a larger magnitude and their centroids 
shift to higher energies 
in the MD case, with respect to the MI one.  This finding can be connected
to the impact of the effective mass on the features of collective modes, 
such as the GDR, 
for which MD interactions lead to
a higher oscillation frequency and to an increase of the Energy Weighted Sum Rule
\cite{zhengPRC2016}.   

We need to mention that similar features have been observed for the other impact parameters, though the magnitude of the power spectrum decreases in less
central events. 
As discussed above, 
we expect the damping rate of the DD oscillations to be affected by particle emission and two-body n-n collisions. 
To get a deeper insight into the latter effect, we have performed
simulations scaling the n-n cross section by a global factor
$f_{cs}$.
We note that microscopic calculations indicate that 
the n-n cs is reduced by in-medium effects \cite{Mac94}. However here, 
for illustrative purposes, we will consider $f_{cs}$ = 0.5 and 2.   
In Fig. \ref{ddcs}, the DD evolution and the corresponding power spectrum are shown, at $b = 2$ fm,
for the asystiff EoS (MI) and the SAMi-J31 (MD) case, 
with the different choices of the n-n cs. 
As one can see, the centroids of the power spectra are almost independent 
on the cs choice. This is observed in both MI and MD cases and confirms that
the DD oscillation frequency 
is mainly determined by the ingredients of the effective interaction. 
The DD signal, with double n-n cross section, dies out faster than the oscillations corresponding to smaller cs values.
Indeed, the larger the cs, the more important  the dissipation effects are
and the dinuclear system gets thermalized in a shorter time. 
Correspondingly, the magnitude of the dipole signal decreases.  


It is interesting to note that 
the main features of the results discussed so far can be simply understood 
in terms of a damped harmonic oscillator.
In this case, indicating by 
$\omega_0$ and
$\tau$ oscillation frequency and damping time, one obtains the following expression for the Fourier transform of the dipole acceleration \cite{baranPRC2009}: 
\begin{equation}
 |D''(\omega)|^2 = \frac{(\omega_0^2+1/\tau^2)^2 D_0^2}{(\omega-\omega_0)^2 + 1/\tau^2}. \label{psana}
\end{equation} 

From the above equation, it emerges in a transparent manner that the strength of
the DD emission is tuned by the 
dipole amplitude $D_0$, but it also reflects the oscillation frequency, being larger
for larger $\omega_0$ values. 
This elucidates the results shown in Fig. \ref{ddeos}, where a larger strength of the spectrum corresponds to a larger value of the centroid. 
On the other hand, as an effect of the 
denominator of Eq. (\ref{psana}), a short damping rate, $1/\tau$, (i.e., a huge number of n-n collisions) leads to a suppression of the strength, as it is shown 
in Fig. \ref{ddcs}.     


We now move to discuss pre-equilibrium nucleon emission.  
Nucleon and light particle emission, happening at the first stages of 
nuclear reactions 
has been widely investigated in 
the literature as a tool to learn about specific properties of the nuclear
effective interaction. In particular, for reactions at Fermi energies, 
it has been shown that the isotopic content can be connected to the behavior
of the symmetry energy at subsaturation density, because the emission mainly
occurs during the expansion phase of the nuclear 
composite system \cite{barPR2005,li08,zhangPLB2015}.

We identify the nucleons emitted 
by looking at the test particles belonging to density
regions with $\rho < 0.01~$fm$^{-3}$, at
our final calculation time, $t_{max}$ = 600 fm/c.  

In Table \ref{tb1}, we report the total number 
and the corresponding N/Z ratio, as obtained for the several 
EoS and cs combinations,  for two impact parameters (b = 2 and 6 fm).  
At b= 6 fm, 
the nucleon emission is generally reduced and 
the N/Z is systematically larger, with respect to b = 2 fm, 
though the effect is not so pronounced.   One can argue that semi-peripheral 
collisions
are less dissipative, thus less nucleons are emitted, with a bigger contribution
from the neutron rich contact region between the two colliding nuclei. 
As a general feature, one also observes that MD interactions are associated
with a larger nucleon emission, as an effect of the larger repulsion induced
by the effective mass.  Larger cs values also produce a more abundant 
pre-equilibrium emission.  On the other hand, the N/Z ratio is smaller in the MD case, 
with respect to the MI one,  and also in calculations employing a larger
n-n cross section.  
One would expect the N/Z ratio to be mainly ruled by Coulomb
effects and by the isovector channel of the effective interaction.
However, when repulsive effects associated with the isoscalar channel (through
the effective mass) and n-n collisions are more important,
the relative weight of isospin effects becomes smaller 
and the N/Z ratio gets closer to 1. 
\begin{table}[tbp]
\begin{tabularx}{0.49\textwidth}{|c|*{2}{>\centering X}|*{2}{>\centering X}|}
\hline
 & \multicolumn{2}{c|} {b = 2 fm} & \multicolumn{2}{c|} {b = 6 fm } \tabularnewline[.01cm] 
\hline
 Interaction & $A_{emit}$ & N/Z  & $A_{emit}$ & N/Z \tabularnewline[.01cm] 
 \hline
asysoft &  17.18 & 2.049  & 16.24  & 2.151 \tabularnewline[.01cm] 
\hline
asystiff & 16.60  & 1.928 & 15.73 & 2.053 \tabularnewline[.01cm] 
\hline
asysuperstiff & 16.22 & 1.774  & 15.38 & 1.870 \tabularnewline[.01cm] 
\hline
SAMi-J27 & 23.50 & 1.433   & 22.29 & 1.489   \tabularnewline[.01cm] 
\hline
SAMi-J31  & 23.06  & 1.571   & 22.04  & 1.631  \tabularnewline[.01cm] 
\hline
SAMi-J35 & 22.90 & 1.687  & 22.03 & 1.741 \tabularnewline[.01cm] 
\hline
asystiff ($f_{cs}$=0.5) & 15.48 & 2.002   & 14.55 & 2.167  \tabularnewline[.01cm] 
\hline
asystiff ($f_{cs}$=2) & 17.28 & 1.878   & 16.54 & 1.972  \tabularnewline[.01cm] 
\hline
SAMi-J31 ($f_{cs}$=0.5) & 20.54 & 1.685  & 19.41  & 1.785  \tabularnewline[.01cm] 
\hline
SAMi-J31 ($f_{cs}$=2) & 24.90 & 1.497  & 23.93  & 1.538 \tabularnewline[.01cm] 
\hline
\end{tabularx}
\caption{The total number of pre-equilibrium emitted nucleons and corresponding N/Z ratios, for the reaction $^{132}Sn+^{58}Ni$ at E/A=10 MeV at b=2 fm and b=6 fm, with different EoS and cs values.}
\label{tb1}
\end{table}

The sensitivity to the symmetry energy emerges when looking at the results
obtained for the three parametrizations employed for each group (MI and
MD) of interactions. Since larger N/Z ratios signal a  larger 
value of the symmetry energy (i.e., a stronger neutron 
repulsion in neutron-rich systems), 
our results indicate that pre-equilibrium nucleon observables mostly
probe a density region in between $\rho_c$ and  $\rho_0$.
Indeed, an opposite 
trend of N/Z with respect to $L$ is observed in the MI and MD cases (see Table \ref{tb1}), reflecting
the different behavior of the symmetry energy, in the density window indicated above, for the two classes of parametrizations (see Fig. \ref{eossym}). 

On the basis of the results discussed so far, one can certainly conclude that the
pre-equilibrium $\gamma$ radiation, as well as the N/Z ratio 
of the emitted particles, is sensitive not only to the symmetry energy
behavior, but also to other aspects, such as the effective mass and the
n-n cs.   
\begin{figure}
\begin{center}
\includegraphics*[scale=0.36]{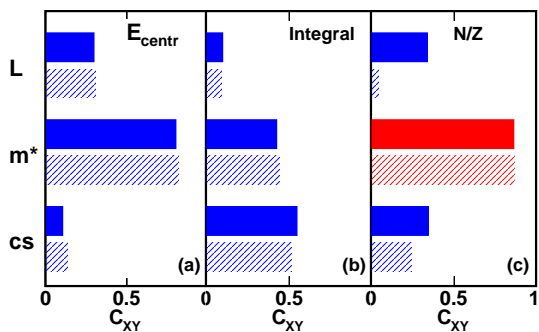}
\end{center}
\caption{(Color online) The correlation functions between the model parameters and selected observables. Solid and shaded bars correspond to different sets
of calculations included in the analysis (see text).  
The blue color indicates a negative correlation,
whereas the red color denotes a positive correlation.}
\label{correlation}
\end{figure} 
In order to show, in a global manner, 
how the model parameters, from EoS and two-body collisions, 
affect the mechanisms considered in our study, a covariance analysis has been adopted to explore 
mutual correlations \cite{zhangPLB2015, irelandJPG2015}. 
The correlation coefficient $C_{XY}$ between the variable $X$ and observable $Y$ is determined as:
\begin{equation}
C_{XY} = \frac{cov(X, Y)}{s(X)s(Y)},
\end{equation}

\begin{equation}
cov(X, Y) = \frac{1}{n-1}\sum_{i=1}^n(X_i-{\bar X})(Y_i-{\bar Y}),
\end{equation}
where $cov(X, Y)$ is the covariance, $s(A)$ and $\bar A$ represent the variance and average of A (=X or Y), respectively. These quantities are evaluated
from the set of $n$ = 10 calculations indicated in Table I (first column).   
 $C_{XY}=\pm 1$ corresponds to  a linear dependence between $X$ and $Y$,
whereas
$C_{XY}=0$ indicates no correlation. Guided by the results discussed above,
we consider three model parameters: the symmetry energy slope $L$, the effective mass and the n-n cs. We also select three
observables, which could be extracted experimentally \cite{pierrPRC2009,zhangPLB2015}: 
the centroid ($E_{centr}$) and the integral of the power spectrum of the DD oscillations, 
and the $N/Z$ ratio of the pre-equilibrium emitted particles. 
In Fig. \ref{correlation}, the solid bars show the correlations between the model parameters and
the observables.  
These results have been obtained by excluding the SAMi-J27 and SAMi-J35 
interactions from the analysis. This is because the SAMi-J parametrizations are characterized by
a (J-$L$) correlation, such that they all give the same symmetry energy value
at $\rho_c$ (see Fig.1).  Since we are interested in the impact of the $L$ parameter
on the observables considered, this correlation could bias the results. 
Indeed, leaving these two interactions out, we are left with a set of EoS which
all have the same $J\approx 30-31$ MeV, and we can pin down
  the sensitivity to the $L$ parameter, 
that determines the behavior of the symmetry energy at subsaturation density. 

However, we also performed the covariance analysis including 
all the calculations listed in Table I, and the corresponding correlation coefficients 
are shown by the shaded bars in Fig. \ref{correlation}. One can see that the results of the two analyses 
are almost the same, except for the correlation between $L$ and N/Z, which is now much
reduced.  
This simply reflects the
opposite behavior observed, in MI vs. MD interactions, for the pre-equilibrium N/Z ratios 
(see Table \ref{tb1} and the above discussion). 
Indeed, the impact of the
derivative of the symmetry energy, $L$,  on results which 
 are sensitive to the
density window between $\rho_c$ and $\rho_0$ is opposite in the SAMi-J, 
with respect 
to the MI interactions, and the global $L$ effect may cancel out when considering
all parametrizations in the covariance analysis. 

The results of the covariance analysis are consistent with all the features 
illustrated above for the MI and MD EoS:  
the correlation functions show, in a more quantitative manner, that the observables considered in our study, apart from the expected sensitivity to the
symmetry energy, are largely influenced by the momentum dependence
of the effective interaction. Moreover, also the n-n cross section has a 
significant impact, comparable to the one associated with the
slope $L$, on the results. 

\section{Conclusions}
To summarize, we have performed a combined study of pre-equilibrium 
dipole radiation and nucleon emission, in low-energy nuclear reactions, 
within a transport model employing
a variety of effective interactions. 
One can conclude that the mechanisms considered, though related 
to isospin effects, are sensitive not only to the 
isovector channel of the interaction, but also to
isoscalar ingredients and to n-n correlations.
In particular, an important sensitivity to the effective mass 
is evidenced. 
For the latter ingredient, however, 
there exist already several theoretical and experimental analyses (in structure and heavy ion 
reactions) of sensitive observables, which may help constraining
its behavior \cite{eff_mass1,eff_mass2}.
Employing several effective interactions in the calculations, 
one can assert that the pre-equilibrium observables considered here probe 
the symmetry energy behavior in the
density window (0.6$\rho_0$ - $\rho_0$), together with the 
corresponding role of n-n correlations.   
Thus, examining 
different observables within the same data set, 
one can hope to constrain at once the details of the n-n cross section
and, 
within the density region investigated in our study,
  the controversial behavior of the symmetry energy. 
An interesting perspective would be to consider, in selected collisions between charge-asymmetric nuclei, the impact
of the ground state deformation and the possible role of pairing effects
on pre-equilibrium dipole oscillations and particle emission, 
to 
further explore the role of structure effects in reaction mechanisms. 



\section{Acknowledgements}


This work for V. Baran was supported by a grant of the Romanian National
Authority for Scientific Research, CNCS - UEFISCDI, project number PN-II-ID-PCE-2011-3-0972.

This project has received funding from the European Union’s Horizon 2020
 research and innovation programme under grant agreement N. 654002.


\end{document}